\documentstyle[aps,prl,multicol,graphicx,epsfig]{revtex}
\begin{document}
\title{Size-dependent effects on electrical contacts to nanotubes and
nanowires}
\author{Fran\c{c}ois L\'{e}onard$^{\ast }$ and A. Alec Talin}
\address{Sandia National Laboratories, MS9161, Livermore, California 94551}
\date{\today }
\maketitle
\draft

\begin{abstract}
Metal-semiconductor contacts play a key role in electronics. Here we show
that for quasi-one-dimensional (Q1D) structures such as nanotubes and nanowires,
side contact with the metal only leads to weak band re-alignement, in
contrast to bulk metal-semiconductor contacts. Schottky barriers are much
reduced compared with the bulk limit, and should facilitate the formation of
good contacts. However, the conventional strategy of heavily doping the
semiconductor to obtain ohmic contacts breaks down as the nanowire diameter
is reduced. The issue of Fermi level pinning is also discussed, and it is
demonstrated that the unique density of states of Q1D
structures makes them less sensitive to this effect. Our results agree with
recent experimental work, and should apply to a broad range of
Q1D materials.
\end{abstract}


\begin{multicols}{2}

The early work of Schottky, Mott and Bardeen has laid the course for much of
the fundamental understanding and improvement in the performance of
electrical contacts to bulk semiconductors. However, as new
nanomaterials are explored for nanoelectronics, the fundamental
aspects of contacts need to be re-examined due to the unique properties of
nanostructures. An example is carbon nanotubes (NTs): despite much
experimental work, it is still unclear whether the contacts are Schottky or
ohmic, with reports of Schottky contacts for Ti\cite{ibm1} and ohmic
contacts for Au\cite{mceuen} and Pd\cite{dai1,dai2}. However, recent
experimental work\cite{chen,kim} has suggested that the type of contact
depends on the NT diameter, with Schottky contacts for small diameter NTs
and ohmic contacs for large diameter NTs.

From a theoretical perspective, it has been demonstrated that Fermi level
pinning (crucial in traditional semiconductors) is ineffective for
quasi-one-dimensional nanostructures {\it end-bonded} to metals\cite{leonard}%
. For NTs {\it side-contacted} by a metal, modeling has been used to extract
Schottky barriers from experimental measurements\cite{chen}, but have not
addressed the origin of the Schottky barriers; and atomistic calculations
have provided case-by-case studies\cite{xue,shan,park,xue2}. However, a more
general theoretical understanding for side-contacts to
Q1D structures is still missing, especially in light of the recent
experimental findings.

In this paper, we present a theoretical and modeling analysis of side
contacts to nanotubes and nanowires. We show that the concepts
developed for bulk metal-semiconductor contacts do not simply carry over to
the nanoscale. In particular, band re-alignement due to charge transfer is
weak due to the limited available depletion width. In NTs, this leads to
relatively small and slowly varying Schottly barriers with NT diameter. In
nanowires (NWs), there is a range of diameters with minimized Schottky
barriers, providing optimal contact properties. We also demonstrate that in
general, Q1D structures are much less sensitive to Fermi level pinning than
their bulk counterparts. Finally, a conventional strategy for making ohmic
contacts is to heavily dope the semiconductor near the contact; we show that
at typical dopings, the contact resistance increases rapidly as the nanowire
diameter is decreased.

We begin by describing the contact geometry considered here. Figure 1a shows
a sketch of a cross section of the contact consisting of a Q1D structure
embedded in a metal. For explicit systems, we consider a single-wall NT, as
shown in Fig. 1b, or a solid nanowire as in Fig. 1c. For the NT, the metal
forms a cylindrical cavity of radius $R+s$ where $R$ is the NT radius and $%
s=0.3$ nm is the distance between the NT and the metal, while for the NW we
consider a solid, continuum cylinder embedded in a perfect metal, with a
sharp interface between the nanowire surface and the metal.

In the simplest picture, the difference between the metal Fermi level $E_{F}$
and the semiconductor valence band edge 

\begin{figure}[h]
\psfig{file=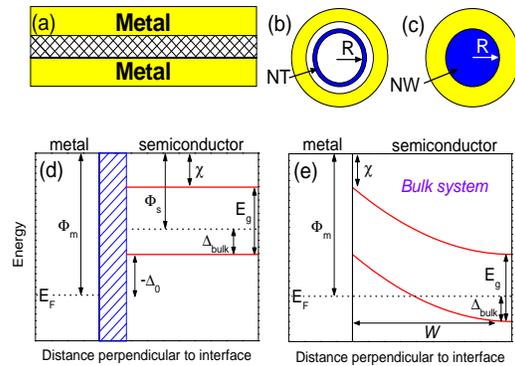,height=137pt,width=194pt}
\caption{Panel (a) shows a cross-section of the contact along the length of
the nanostructure. Panels (b) and (c) show radial cross-sections for
metal-nanotube and metal-nanowire contacts. Panel (d) shows the band
alignment before charge transfer. In a bulk contact, panel (e), band bending
over a distance $W$ leads to a Schottky barrier $\Delta _{bulk\text{.}}$}
\end{figure}

$E_{v}$ (the barrier for holes) is
 (Fig. 1d)%
\begin{equation}
\Delta _{0}=E_{g}+\chi -\Phi _{m}  \label{bare}
\end{equation}%
where $\Phi _{m}$ is the metal workfunction, $\chi $ is the semiconductor
electron affinity and $E_{g}$ is the semiconductor band gap. A positive
value for $\Delta _{0}$ indicates a Schottky barrier, while a negative value
indicates an ohmic contact. Because bandgap decreases with increasing
diameter for Q1D structures, the value of $\Delta _{0}$ depends on the
nanostructure diameter. The behavior of Eq. $\left( \ref{bare}\right) $ for
undoped NTs is shown in Fig. 2 for a value of $\Phi _{m}-\Phi _{NT}=0.4$ eV
(typical of Pd), and using the relation $E_{g}=2a\gamma /d$ between bandgap
and NT diameter $d$ ($a=0.142$ nm is the C-C bond length, $\gamma =2.5$ eV
is the tight-binding overlap integral, and $\Phi _{NT}$ is the NT
workfunction assumed to be at midgap for an undoped NT). One problem with
this picture (besides the fact that the physics is incomplete, as will be
discussed below) is that Eq. $\left( \ref{bare}\right) $ predicts large and
negative values for $\Delta _{0},$ signaling strong ohmic contacts. However,
it is clear that such strong ohmic contacts are not observed experimentally.

In general, charge transfer between the metal and semiconductor leads to
band re-alignement. At a bulk semiconductor junction (Fig. 1e) this charge
transfer leads to the Schottky barrier%
\begin{equation}
\Delta _{bulk}=E_{g}+\chi -\Phi _{s}  \label{bulk}
\end{equation}%
where $\Phi _{s}$ is the semiconductor workfunction. This relationship
arises because, in the bulk system, a depletion width $W$ perpendicular to
the metal-semiconductor interface is created until the band lineup in Eq. $%
\left( \ref{bulk}\right) $ is obtained. However for Q1D structures, the
depletion width depends {\it exponentially} on the doping\cite{leonard2} and
is much longer than the device size for non-degenerate doping, leading to
slowly varying bands outside of the contact; and for a three-terminal device
the band-bending in the channel is governed by the gate voltage. In either
case, the band alignment

\begin{figure}[h]
\psfig{file=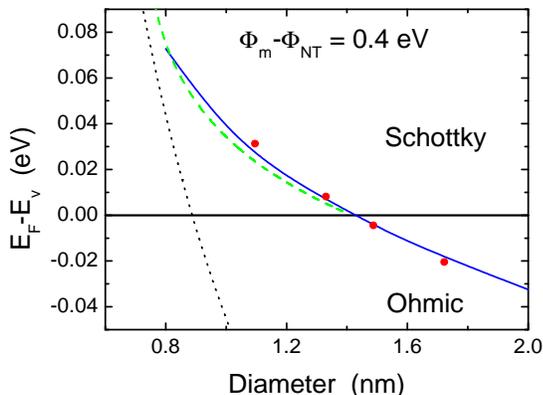,height=153pt,width=204pt}
\caption{Schottky barrier at nanotube-metal contacts for parameters typical
of Pd. Dotted line is from Eq. (\ref{bare}), solid line is
 from Eqs (\ref{sigma}) and (\ref{potential}), dashed line is Eq. (%
\ref{delta}) and circles are calculated from an atomistic approach.}
\end{figure}

is determined by that in the contact. But for a
side-contacted Q1D structure, the semiconductor is only a few nanometers
thick in the direction perpendicular to the metal-semiconductor interface;
thus only a region of the order of the nanostructure cross-section can be
depleted, giving partial band re-alignement. The value of $\Delta $ will
then be somewhere between $\Delta _{0}$ and $\Delta _{bulk}$ (for an undoped
NT or NW, $\Delta _{bulk}=E_{g}/2,$ which would always give relatively high
Schottky barriers).

Nanotubes are an extreme example of this situation, since the possible
``depletion width'' is the size of the NT wall; the charge transfer and
image charge in the metal create two nested hollow cylinders with opposite
charge, and an associated electrostatic potential. The charge and potential
must be self-consistent. We can capture this behavior using
analytical models for the charge and potential. The charge per unit area on
the NT is expressed as%
\begin{equation}
\sigma =eN\int D_{NT}(E+eV_{NT})f(E-E_{F})dE  \label{sigma}
\end{equation}%
where $D_{NT}(E)$ is the NT density of states\cite{white} shifted by the
electrostatic potential on the NT, $f\left( E-E_{F}\right) $ is the Fermi
function, and $N=4/(3\sqrt{3}a^{2})$ is the atomic areal density. We assume
a uniform and sharp distribution of the charge on the NT, and all
calculations presented in this paper are for room temperature.

For the geometry of Fig. 1, solution of Poisson's equation gives the
potential on the NT as%
\begin{equation}
eV_{NT}=-\sigma \frac{eR}{\varepsilon _{0}}\ln \frac{R+s}{R}=-\frac{e^{2}}{C}%
\sigma  \label{potential}
\end{equation}%
where $\varepsilon _{0}$ is the permittivity of free space and $C$ is the
capacitance per unit area between the metal and the NT. Equations $\left( %
\ref{sigma}\right) $ and $\left( \ref{potential}\right) $ can be solved
self-consistently for a given NT. In this model the electrostatic potential
induced on the NT modifies the barrier to $\Delta =\Delta _{0}-eV_{NT}$.
Figure 2 shows results of such calculations for parameters typical of Pd.
Clearly, the behavior is different from the simple expressions in Eqs $%
\left( \ref{bare}\right) $ and $\left( \ref{bulk}\right) $. Indeed, the bulk
limit $\Delta _{bulk}=E_{g}/2$ gives very large barriers, much too large to
even appear on the scale of Fig. 2. The results suggest that there is a
transition between Schottky and ohmic behavior at a NT diameter around 1.4
nm, in agreement with recent experimental data for Pd contacts%
\cite{chen,kim}. We have verified these predictions using an atomistic
description of the NT based on a self-consistent, tight-binding Green's
function formalism. As shown in Fig. 2, the results of such calculations
indicate excellent agreement with the analytical approach introduced above.

To proceed further we focus on the small and positive $\Delta $ regime;
approximation of the integral in Eq. \ref{sigma} leads to 
\begin{equation}
\sigma =\frac{eNa\sqrt{3}}{2\sqrt{2\beta }\pi ^{3/2}R\gamma }\sqrt{\frac{%
E_{g}kT}{2}}e^{-\beta \frac{\Delta }{kT}}  \label{chargeanalytical}
\end{equation}%
with $\beta =0.7$. Combined with Eq $\left( \ref{potential}\right) $ this
gives the Schottky barrier%
\begin{equation}
\Delta \approx \frac{kT}{\beta }\ln \left( \frac{\alpha \sqrt{\frac{E_{g}}{%
2kT}}}{\ln \alpha \sqrt{\frac{E_{g}}{2kT}}-\Delta _{0}/kT}\right)
\label{delta}
\end{equation}%
where $\alpha =\left( e^{2}Na\sqrt{3}\right) /\left( 2\sqrt{2\beta }\pi
^{3/2}R\gamma C\right) $. The behavior of this function is plotted in Fig.
2, showing good agreement with the full calculation. The logarithmic
dependence implies relatively slowly varying $\Delta ,$ at least compared
with Eq. $\left( \ref{bare}\right) $. The NT diameter delimiting Schottky
from ohmic behavior is\cite{remark} 
\begin{equation}
d\approx d_{0}\left( 1+\alpha \sqrt{\frac{kT}{\Phi _{m}-\Phi _{NT}}}\right) ,
\end{equation}%
where $d_{0}$ is the crossover diameter that would be obtained from Eq. $%
\left( \ref{bare}\right) $. Thus the crossover diameter is increased 
by $\delta d=\alpha \sqrt{\frac{kT}{\Phi _{m}-\Phi _{NT}}}%
d_{0}$. Making ohmic contact to a wide range of NT diameters requires a
small $\delta d$; this can be accomplished at low temperature, with a large
metal workfunction, or with a large capacitance (giving a small $\alpha $).
Embedded contacts thus provide an advantage over planar contacts because of
their larger capacitance.

We now consider side-contacts to nanowires, where the possible depletion
width increases with diameter, and the dependence of the bandgap on diameter
is different than in NTs. We model a NW with density of states%
\begin{equation}
D_{NW}(E)=\frac{\sqrt{2m^{\ast }}}{\pi \hbar }\left( E-E_{g}/2\right) ^{-1/2}
\label{dosNW}
\end{equation}%
where $m^{\ast }$ is the effective mass. For silicon NWs, it has been shown
experimentally\cite{SiNW} that the band gap depends on diameter as $%
E_{g}=E_{0}+C/d^{2}$ where $E_{0}=1.12$ eV and $C=4.33$ eVnm$^{2}$. We
consider the situation $\Phi _{m}-\Phi _{NW}=0.7$ eV typical of contacts to
Si. Fig. 3a shows the expected Schottky barrier heights from Eq. $\left( \ref%
{bare}\right) ,$ which predicts ohmic contacts to NWs with diameters larger
than 4 nm. To study the effects of charge transfer, we perform a
self-consistent calculation of the charge and potential, using Eq. $\left( %
\ref{dosNW}\right) $ to obtain the charge and solving Poisson's equation
numerically in the NW to obtain the potential (we use an atomic volume
density $N_{v}=5\times 10^{28}$ atoms/m$^{3}$).

Fig. 3b,c shows the self-consistent band-bending for NWs of 2 and 10 nm
radius. Clearly, the nanoscale dimension of the NWs prevents the bands from
reaching their asymptotic value; instead, there is only a weak band-bending
present. To quantify the Schottky barrier height, we calculate the spatial
average of $E_{F}-E_{v}(r)$; the results plotted in Fig. 3a indicate that
the 

\begin{figure}[h]
\psfig{file=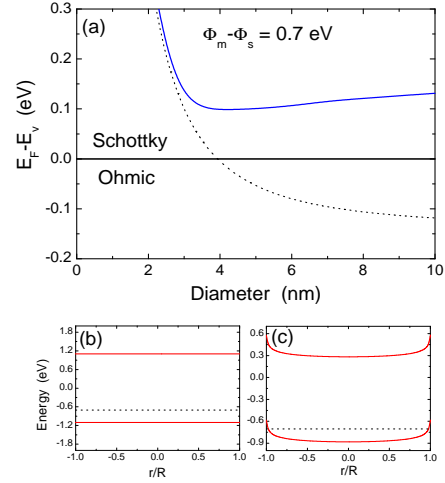,height=185pt,width=166pt}
\caption{Panel (a): Schottky barrier at nanowire-metal contacts for
parameters typical of SiNWs. Dotted line is from Eq. (\ref{bare})
and solid line is self-consistent calculation. Panels (b) and (c):
Band-bending across nanowires with diameters of 2 nm and 10 nm,
respectively; dotted lines are Fermi level.}
\end{figure}

contact is always of Schottky character, with the barrier minimized at a
diameter of about 4 nm. Thus, while in NTs the barrier height decreases
monotonically with diameter, the behavior in other Q1D structures may be
non-monotonic, with a range of diameters providing optimal contact
properties. We also note that, just as for NTs, the barrier heights are much
smaller than the bulk limit $\Delta _{bulk}=E_{g}/2$ (not shown in Fig. 3
for clarity).

In a bulk metal-semiconductor contact, metal-induced gap states (MIGS) lead
to Fermi level pinning, and modification of the Schottky barrier height to $%
\Delta _{pin}$\cite{leonard}. To model this effect in side contacts to Q1D
structures, we consider a radial pinning charge

\begin{equation}
\sigma _{pin}(r)=D_{0}N_{A}\left[ E_{F}-E_{N}(r)\right] h(r)
\end{equation}%
where the neutrality level $E_{N}$ is at midgap [i.e. $E_{N}(r)=-eV(r)$], $%
h(r)=e^{-r/l}$ for a NW and $h(r)=\delta _{r,R}$ for a NT, and $N_{A}=N$ for
a NT and $N_{A}=N_{v}^{2/3}$ for a NW. We choose $l=0.3$ nm, a typical value
for metal-semiconductor interfaces\cite{louie}. We add this pinning charge
to Eq. $\left( \ref{sigma}\right) $ or to the charge calculated from Eq. $%
\left( \ref{dosNW}\right) $ and repeat our self-consistent calculations.

Figure 4a shows the Schottky barrier calculated for several NTs as a
function of the density of gap states ($\Delta _{pin}=E_{g}/2$). Clearly,
there is a rapid onset of pinning at $D_{0}\sim 0.1$ states/(atom$\cdot $%
eV); this value of $D_{0}$ is rather large considering the van der Waals
bonding of NTs to metals, and atomistic calculations\cite{xue,park} have
obtained seemingly small values. Thus, as in end-bonded contacts, we expect
that Fermi level pinning will play a minor role in side-contacts to NTs.

\begin{figure}[h]
\psfig{file=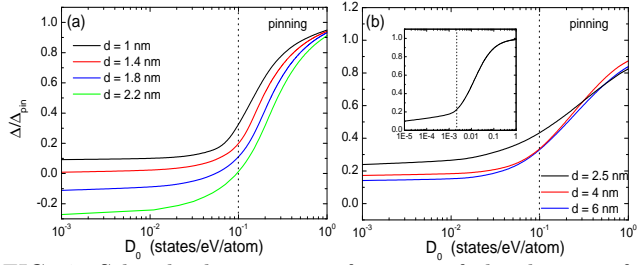,height=100pt,width=240pt}
\caption{Schottky barrier as a function of the density of gap
states for several NTs (a) and NWs (b). The inset in (b) shows the behavior
for a planar metal-semiconductor contact. }
\end{figure}

Figure 4b shows the effects of Fermi level pinning on the barrier height in
SiNWs. The results also indicate a value of $D_{0}\sim 0.1$ states/(atom$%
\cdot $eV) required to see pinning effects. For comparison, the inset in
this figure shows the same calculation for a bulk metal-semiconductor
interface with the same parameters, indicating that only $0.002$ states/(atom%
$\cdot $eV) are needed to reach the onset of pinning. Thus, the Q1D system
requires almost two orders of magnitude larger density of pinning states
compared with the bulk interface.

The origin of this behavior can be traced to the unique density of states of
Q1D systems. Indeed, for Si, we can repeat the analysis leading to Eq. $%
\left( \ref{chargeanalytical}\right) $ using the density of states for the
NW and for the bulk system [$D_{bulk}(E)=\sqrt{2}\left( m^{\ast }\right)
^{3/2}\left( \pi ^{2}\hbar ^{3}\right) ^{-1}\sqrt{E-E_{g}/2}$]. This leads
to the ratio $\sigma _{NW}/\sigma _{bulk}=\left( 2\pi N_{v}^{1/3}\beta
\right) /\left( m^{\ast }kT\right) $. The appearance of the $kT$ factor in
the denominator is entirely due to the Q1D density of states of the NW and
the presence of a van Hove singularity at the band edge. At room
temperature, we find that $\sigma _{NW}/\sigma _{bulk}>100$; thus the MIGS
are competing with a much larger charge density in the Q1D system.

Our discussion has so far focused on the situation of low doping, where the
strategy for making ohmic contacts is by selection of a metal with
appropriate workfunction. In traditional metal-semiconductor contacts, an
alternative approach is to heavily dope the semiconductor, and rely on
tunneling through the Schottky barrier to reduce the contact resistance and
obtain ohmic-like contacts. To address the feasibility of this approach for
contacts to NWs, we repeat our self-consistent calculations for the Si NW,
focusing on the situation where the metal Fermi level is in the middle of
the NW bandgap at the interface, and adding a uniform doping charge of $%
10^{19}e/cm^{-3}$. Figure 5 shows the band-bending in the presence of this
doping charge for NWs of 40 and 10 nm diameters. We calculate the contact
conductance from%
\begin{equation}
G\sim \int_{E_{c}^{\min }}^{\infty }T(E)\left( -\frac{\partial f}{\partial E}%
\right) dE
\end{equation}%
where the tunneling probability $T(E)$ is obtained from the

\begin{figure}[h]
\psfig{file=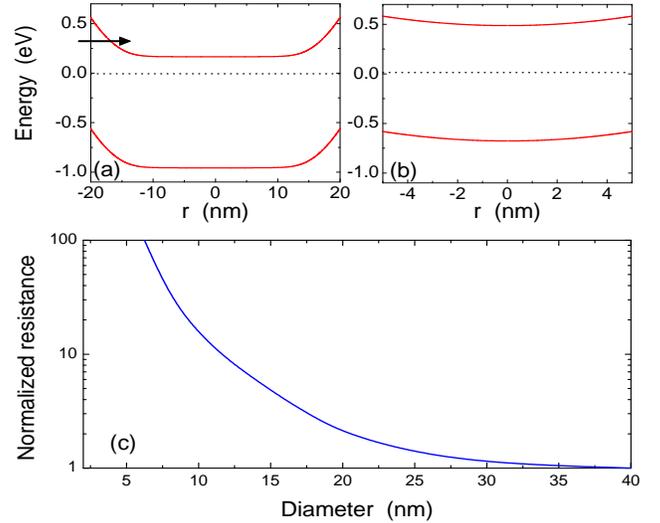,height=200pt,width=240pt}
\caption{Band-bending across Si NWs with doping of 10$^{19}e/cm^{-3}$ for
diameters of 40 nm (a) and 10 nm (b). The arrow indicates tunneling of
electrons throught the Schottky barrier. The normalized resistance is shown
in figure (c) as a function of NW diameter.}
\end{figure}

WKB approximation. The normalized contact resistance is then $G_{\infty }/G$
where $G_{\infty }$ is the conductance in the limit of large diameters. The
behavior of the normalized resistance as a function of NW diameter is shown
in Fig. 5c, indicating a rapid increase of the resistance with decrease in
diameter. The origin of this behavior is the increased tunneling
distance and reduced range of tunneling energies because of the poor
band-bending. One implication of this result is that different
diameter NWs will require different doping levels to achieve the same
contact quality.

In summary, we find that the concepts developed to describe traditional
metal-semiconductor interfaces fail to properly account for the properties
of contacts to Q1D structures. Optimizing device performance will not only
require selecting Q1D structures for their behavior in the channel, but also
for their contact properties. We expect that our results will be applicable
to a broad range of Q1D structures.

Sandia is a multiprogram laboratory operated by Sandia Corporation, a
Lockheed Martin Company, for the United States Department of Energy under
contract DE-AC01-94-AL85000.

$^{\ast }$email:fleonar@sandia.gov

\end{multicols}


\begin{references}
\bibitem{remark} To derive this relation we assumed a constant capacitance
over the range of NT diameters of interest. A more detailed treatment would
not change the key physics.
\end{references}
\end{document}